\documentclass{article}
\begin{document}

\title {Effect of Massive Fields on Inflation}

\author{S.G. Rubin \\Moscow Engineering Physics Institute, Moscow,
Russia;
\\Center for CosmoParticle Physics"Cosmion";
\\e-mail: serg.rubin@mtu-net.ru}

\maketitle

\begin{abstract}
Effects caused by an additional massive scalar field interacting
with an inflaton field are analyzed. Inflation is shown to have two
stages, the first of which is dominant and characterized by
ultraslow dynamics of the inflaton field. Constraints on the model
parameters are obtained.\end{abstract}

The occurrence of inflationary era in the universe evolution seems
to be inevitable because it allows the explanation of a great
number of observed facts \cite{Linde90,Khlopov}. Early inflation
mechanisms \cite{Star80,Guth81} were based on the consistent
equations of scalar and gravitational fields. Nevertheless, the
simplest inflation models could not explain the totality of
observed data. In particular, the predictions of the chaotic
inflation model [5, 6] about temperature fluctuations in cosmic
background radiation do not contradict observations only for a
rather unnatural form of the inflaton field potential (see also
[7]).

At the same time, the interaction of a large number of various
fields existing in nature should give rise to new phenomena in
inflation scenario. Further development of the theory has led to
the emergence of inflation models involving additional fields,
among which are the models of hybrid inflation \cite{Linde91a} and
inflation on the pseudo Nambu  - Goldstone field \cite{Freese94}.
The interaction of the classical inflaton field with other
particles produced by it is one of the basic elements of some
inflation models. This effect provides a basis for the warm
inflation scenario \cite{Berera00}, which, however, is not free
from flaws \cite{Yoko98a}; back reaction of the produced particles
on the dynamics of inflaton field was considered in \cite{Dolgov98}
è \cite{Dymnikova00}.

The purpose of this work is to study the back reaction of an
additional field on the classical motion of the basic inflaton
field. It is assumed that the additional field is massive enough
for it to be at the minimum of its effective potential during
inflation. Nevertheless, it is shown below that its influence can
noticeably decelerate the system motion.

In what follows, the simplest form of interaction is considered
allowing the analytical results to be obtained. Namely, we
introduce, apart from the inflaton field $\varphi$, an additional
scalar field $\chi$ and write the action in the form
\begin{equation}\label{mi1}
S=\int d^4x\sqrt{-g}\left[ \frac 12\varphi _{,\mu } \varphi ^{,\mu
}-V(\varphi )+\frac 12\chi _{,\mu }\chi ^{,\mu }- \frac 12m^2\chi
^2-\kappa \chi u(\varphi )\right] .
\end{equation}
where $u(\varphi)$ is a polynomial of degree no higher than three
for the renormalizable theories. Below, $u(\varphi)=\varphi ^2$ is
taken for definiteness. The first power of the field $\chi$ in the
interaction is necessary in order to obtain compact analytical
results valid for an arbitrary coupling constant $\kappa$, rather
than the expansion in powers of this constant. The interaction of
this type arises in supersymmetric theories and is considered in
hybrid inflation scenarios \cite{Lyth96}. Dolgov and Hansen
\cite{Dolgov98} used this type of interaction in studying the back
reaction of produced particles on the motion of classical field.

The set of the classical equations for both fields is written as
\begin{equation}\label{ad1}
\begin{array}{l}
\frac{1}{{\sqrt { - g} }}\partial _\mu  \left( {\sqrt { - g}
\partial ^\mu  \chi } \right) + m_\chi ^2 \chi  + \kappa \varphi ^2
= 0, \\
\\
\frac{1}{{\sqrt { - g} }}\partial _\mu  \left( {\sqrt { - g}
\partial ^\mu  \varphi } \right) + V'(\varphi ) + 2\kappa \varphi
\chi  = 0. \\
\end{array}
\end{equation}
Let us consider the case of heavy $\chi$ particles. In the
inflationary era, this means that
\begin{equation}\label{ad2}
m_{\chi} >> H(\varphi ),
\end{equation}
and the Hubble constant $H(\varphi )$ is determined by the slowly
varying classical field $\varphi $. The first of Eqs.(\ref{ad1})
can be brought to the form
\begin{equation}\label{ad3}
\chi (x) =  - \kappa \int {G(x,x')\varphi ^2 (x')dx'}.
\end{equation}
The right-hand side of Eq.(\ref{ad3}) can be simplified using the
equation for the Green function $G(x, x')$ \cite{Birrell} written
as
\begin{equation}\label{mi4}
G(x,x^{\prime })=\frac 1{m^2}\delta (x-x^{\prime })- \frac
1{m^2}\frac 1{\sqrt{-g}}\partial _\mu \sqrt{-g}
\partial ^\mu G(x,x^{\prime }).
\end{equation}
After two iterations, the field $\chi$ takes the explicit form
\begin{equation}\label{ad4}
\chi (x) \simeq  - \frac{\kappa }{{m_\chi ^2 }}\varphi ^2 (x) +
\frac{\kappa }{{m_\chi ^4 }}\partial ^\mu  \sqrt { - g} \partial
_\mu  \left( {\frac{1}{{\sqrt { - g} }}\varphi ^2 (x)} \right) ,
\end{equation}
which is valid if the derivatives of the inflaton field $\varphi $
are small. Substituting this expression into the second of Eqs.
(\ref{ad1}), one arrives at the following classical equation for
the inflaton field:
\begin{equation}\label{mi6}
\partial _\mu \sqrt{-g}\partial ^\mu \varphi +
\sqrt{-g}V_{ren}^{\prime }(\varphi )+{\frac{2\alpha
^2}{m_{\chi}^2}} \varphi \partial _\mu \sqrt{-g}
\partial ^\mu \varphi ^2 =0,
\end{equation}
where $\alpha \equiv \frac{\kappa}{m_{\chi}}$ a dimensionless
parameter and$V_{ren}(\varphi )=V(\varphi)-\frac{\alpha
^2}{2}\varphi ^4$ is the potential of inflaton field renormalized
due to interaction with the field $\chi$. The last term on the
left-hand side of Eq.(\ref{mi6}) is usually treated as a back
reaction of radiation \cite{Dolgov98}. Equation (\ref{mi6})
corresponds to the effective action for inflaton field

\begin{equation}\label{mi5}
S_{eff}=\int d^4x\sqrt{-g}\left[ \frac 12\varphi _{,\mu } \varphi
^{,\mu }-V_{ren}(\varphi )-\frac 1{\sqrt{-g}} \frac{\alpha
^2}{2m^2}\varphi ^2\partial _\mu \sqrt{-g}
\partial ^\mu \varphi ^2 \right] .
\end{equation}

Note that the correction $\delta V = -\frac{\alpha ^2}{2}\varphi
^4$ to the potential follows from the analysis of classical Eqs.
(\ref{ad1}). At the same time, the same expression can be obtained
by calculating the first quantum correction to the $\varphi$-field
potential interacting with the field  $\chi$ at zero 4-momenta of
external lines corresponding to the $\varphi$-field quanta. The
internal line corresponds to the $\chi$-field propagator in the $s$
and $t$ channels.

The last term in Eq.(\ref{mi6}) is important for further
consideration. Nonminimal kinetic term arises in equations for
density fluctuations in early universe \cite{Lukash80}. Morris
\cite{Morris01} showed that a change in the form of kinetic term in
the scalar$–$tensor theory leads to the inflation on a lower, than
ordinary, energy scale, in agreement with the conclusions of this
work. Similar result can be obtained by introducing a nonminimal
interaction between an inflaton and a gravitational field [17, 18].

In general, the renormalized potential contains the sum of
contributions from the corrections due to interaction with all
existing fields. In the first model of chaotic inflation with the
$\lambda \varphi^4$ potential, the observed data led to a value of
$\lambda (\sim 10^{-13})$. This means that the corrections
introduced to the expression for by all fields, including the
correction  $\delta V=-(\alpha^2 /2)\varphi^4$ considered in this
work, must cancel with a high accuracy.

Below, it is demonstrated that the renormalization of the kinetic
term allows one, in particular, to weaken significantly the
conditions imposed by the observations on the parameters of the
theory. In weak fields, the contribution from the last term in
Eqs.(\ref{mi6},\ref{mi5}) is negligible. As to the inflation stage,
it can be substantial at large field magnitudes.

During inflation, the field is assumed to be uniform; i.e.,
$\varphi =\varphi (t)$, and Eq.(\ref{mi6}) is greatly simplified.
Taking into account that the scale factor $a$ is expressed in terms
of the Hubble constant $H$ in the ordinary way, $a=exp(\int Hdt)$,
Eq.(\ref{mi6}) can be rewritten as
\begin{equation}\label{mi8}
\frac{d^2\varphi }{dt^2}+3H\frac{d\varphi }{dt}+V_{ren}^{\prime
}(\varphi )+{\frac{4\alpha ^2}{m_{\chi}^2}}\left[ 3H\varphi
^2\frac{d\varphi }{dt}+\varphi ^2\frac{d^2\varphi }{dt^2}+ \varphi
\left( \frac{d\varphi }{dt}\right) ^2\right] =0. \nonumber
\end{equation}
Slow time variation of the field $\varphi$ implies that the terms
proportional to $d^2 \varphi /dt^2$ and $(d\varphi /dt)^2$ are
small. Neglecting them, one obtains the easily integrable equation
\begin{equation}\label{mi9}
\left( {3H + \frac{12H\alpha ^2}{m_{\chi}^2 }}\varphi ^2 \right)
\dot \varphi + V_{ren}'(\varphi ) = 0 .
\end{equation}

In what follows, the nonrenormalized potential is taken in the form
$V(\varphi )=\lambda_0 \varphi ^4$, and, therefore,$V_{ren}=\lambda
\varphi ^4$ where, $\lambda =\lambda_0 -\alpha^2 /2$. Taking into
account the usual relation $H=\sqrt{8\pi V_{ren}(\varphi )/3}/M_P$
between the Hubble constant and the potential, one can easily
obtain the field variable $\varphi$ as an implicit function of
time:
\begin{equation}\label{mi10}
t=\frac{\sqrt{3\pi /2}}{M_{P}\sqrt{\lambda }} \left[\ln (\varphi _0
/\varphi )+ \frac{2\alpha^2}{m_{\chi}^2}( \varphi _0^2- \varphi ^2
) \right] .\end{equation}
Here, the first term reproduces the result of the standard
inflation model. The second term results from the interaction of
the inflaton field and the field $\chi$. It follows from
Eq.(\ref{mi9}) that the second term dominates at
\begin{equation}\label{mi11}
\varphi \geq\varphi _c  \equiv \frac{m_{\chi}}{{2\alpha }}.
\end{equation}
Therefore, there are two inflation stages: the ordinary stage at
$\varphi \leq\varphi _c$ and the ultraslow stage at $\varphi
\geq\varphi _c$. Indeed, the field motion velocity obtained from
(\ref{mi9}) with allowance made for Eq.(\ref{mi11}) is much smaller
than its ordinary value $\dot \varphi = V'/3H$. The first inflation
stage is completed when condition (\ref{mi11})) ceases to be true.
Then the ordinary inflation stage ${\ddot \varphi }<<3H{\dot
\varphi}$ begins and continues as long as the condition is
satisfied.

Because the second stage has been much studied, I will analyze the
first stage, for which the second term in Eqs.
(\ref{mi9},\ref{mi10}) dominates, i.e., for $\varphi >\varphi_c$.
In this case, the field depends on time as
\begin{equation}\label{mi12}
\varphi (t)=\sqrt{\varphi _0^2 - t\frac{M_P m_{\chi}^2}{\alpha ^2
\sqrt{6\pi}}}.
\end{equation}
This expression is derived under the "ultraslow roll-down"
condition, which, according to Eq.(\ref{mi8}), has a rather unusual
form ${\ddot \varphi }<< 12H\varphi ^2 {\dot \varphi
}{\frac{\alpha^2}{m_{\chi}^2}}$.

Let us determine the amplitude of quantum fluctuations arising at
the first inflation stage for the potential $\lambda \varphi ^4$.
This can most easily be done by taking into account that the first
term in Eqs.(\ref{mi6}) and (\ref{mi9}) is much smaller than the
third one and introducing an auxiliary field $\widetilde{\varphi}$,
after which the substitution $\widetilde{\varphi}=(\alpha
/m_{\chi})\varphi^2$ brings action to the form
\begin{equation}\label{ad6}
S = \int {dx\sqrt { - g} \left[ {\frac{1}{2}\partial _\mu  \tilde
\varphi \partial ^\mu  \tilde \varphi  - \frac{1}{2}\tilde m^2
\tilde \varphi ^2 } \right]},
\end{equation}
corresponding to a free massive field with mass $\tilde m \equiv
m_{\chi}\sqrt{2\lambda }/\alpha$. This substitution is valid at the
inflation stage under consideration, when the field value is
positive. The fluctuation amplitude for the massive noninteracting
field is known to be $\Delta \widetilde{\varphi}
=\sqrt{3/(8\pi^2)}H^2 /\tilde m$ \cite{Bunch78}. On the scale of
modern horizon, the constraint on the mass of quanta of this field
is also known: $\tilde m\sim 10^{-6}M_P$, as is obtained from the
comparison with the COBE measurements of the energy-density
fluctuations, $\delta \rho /\rho \approx 6\cdot 10^{-5}$
\cite{COBE}. Expressing $\tilde m$ in terms of the initial
parameters, one obtains the following relation between them:
\begin{equation}\label{ad8}
\frac{{m_\chi  }}{{M_P }}\frac{{\sqrt \lambda  }}{\alpha } \sim
10^{ - 6}
\end{equation}

Let us determine the field $\varphi_U$ at which a causally
connected area was formed, which generated the visible part of the
universe. The number of e-foldings necessary to explain the
observed data is $N_U\approx 60$. Then, using the relation$N_U
=\int _{\varphi_{U}}^{\varphi_{end}}Hdt$

\begin{equation}\label{mi15}
\begin{array}{l}
N_U  = \int\limits_{\varphi _U }^{\varphi _c } {\frac{{H(\varphi
)}}{{\dot \varphi }}} d\varphi  + \int\limits_{\varphi _c
}^{\varphi _{end} } {\frac{{H(\varphi )}}{{\dot \varphi }}}
d\varphi  =  \\ = \frac{{2\pi \alpha ^2 }}{{M_P^2 m_{\chi}^2
}}\left( {\varphi _{_U }^4  - \varphi _c^4 } \right) + \frac{\pi
}{{M_P^2 }}\left( {\varphi _c^2  - \varphi _{end}^2 } \right). \\
 \end{array}
\end{equation}
where it is taken into account that the time dependencies of the
field $\varphi$ at the first and the second inflation stages are
different. The second stage is completed at $\varphi=
\varphi_{end}$. Assuming that the first term containing the initial
value of the field $\varphi_U$ dominates, one obtains the desired
expression
\begin{equation}\label{mi16}
\varphi _U  \simeq \left( {\frac{{N_U }}{{2\pi }}} \right)^{1/4}
\sqrt {\frac{{M_P m_{\chi}}}{\alpha }}.
\end{equation}
Note that the visible part of the universe in this case can be
formed at $\varphi < M_P$, i.e., rather late. This is explained by
the fact that at the first stage the field moves ultraslowly and
the universe has had time to expand to the suitable size.

Expression (\ref{mi16}) differs substantially from the standard
result $\varphi_U \sim M_P$, which is obtained for the inflaton
field with potential $\lambda \varphi ^4$ without regard for the
interaction with the massive fields of other sorts $\varphi _U  =
\sqrt {N_U /\pi } M_P.$

The second term in Eq.(\ref{mi15}) determines the number $N_2$ of
e-foldings at the second inflation stage. Assuming that
$\varphi_c^2 >> \varphi_{end}^2$ and substituting the value
$\varphi_c$ from Eq.(\ref{mi11})), one has
\begin{equation}\label{ad10}
N_2  = \frac{\pi }{4}\left( {\frac{{m_\chi  }}{{\alpha M_P }}}
\right)^2 .
\end{equation}
Evidently, over a wide range of parameters $\alpha$ and $m_{\chi}$,
the second stage may be short or absent at all.

The above arguments are valid if the field $m_{\chi}$ is massive
enough so that it is placed at the minimum of its effective
potential during inflation. As is known, the field rapidly rolls
down to the minimum if the Hubble constant becomes smaller that the
field mass, i.e., if $H < m_{\chi}$. The Hubble constant depends on
time. For this reason, the necessary estimates will be made for the
instant the visible universe originated ($\varphi =\varphi_U$),
when the largest scale fluctuations arise. Simple mathematics gives

\begin{equation}\label{mi17}
m_{\chi} > H(\varphi _U )\quad \to \quad \frac{{\sqrt \lambda
}}{\alpha }\leq \sqrt{\frac{3}{4N_U}} \sim 0.1 .
\end{equation}
This restriction indicates that one cannot fully avoid the fine
fitting of parameters because $\lambda = \lambda_0 - \alpha^2 /2$
and, according to constraint (\ref{mi17}), $\alpha^2 \geq
100\lambda$. Nevertheless, this fitting is weaker than that
requiring the cancellation of all quantum corrections down to a
value of $\sim 10^{-13}$ in the early inflation models with the
potential $\lambda \varphi^4$. Using Eqs.(\ref{ad8}) and
(\ref{mi17}), one can easily obtain a rather weak limitation:
$m_{\chi}\geq 10^{-5}M_P$ on the mass of the additional field
$\chi$.

Thus, a particular example was taken in this work to demonstrate
that massive fields, even being at their minimum (which depends on
the magnitude of inflaton field), can materially decelerate the
motion of the main — inflaton field at the first inflation stage.
Due to the first, ultraslow, stage, the visible universe could form
at $\varphi < M_P$. The second stage precedes the completion of
inflation and evolves in the ordinary way, but is rather short. In
particular, for the parameters $m_{\chi} = 10^{-3}M_P$ and $\lambda
= 10^{–6}$, one has: the visible universe formed at $\varphi_U
\approx 5\cdot 10^{–2}M_P$ ; the first and second stages are
separated at $\varphi_c \approx 5\cdot 10^{–4}M_P$ ; and the second
inflation stage is much shorter than the first one.

The inclusion of the interaction between the inflaton field and
more massive fields enables one to materially weaken the
constraints imposed on the potential parameters by the smallness of
energy density fluctuations, although one fails to fully avoid the
fine tuning of the parameters. The effects considered are
associated with the renormalization of the kinetic term for the
inflaton field interacting with an additional massive field.
Because the similar renormalization takes place for every sorts of
additional fields \cite{Itzykson}, the inclusion of new fields will
enhance the effect of deceleration of classical motion at high
energies.

I am grateful to A.A. Sakharov for useful comments, M.Yu. Khlopov
for interest in the work, and the referee for constructive
criticism. This work was supported by the Russian State Scientific
Engineering Program 'Astronomy. Fundamental Cosmic Research'
(project 'Cosmoparticle Physics') and by the International
Collaboration Cosmion-ETHZ and Epcos-AMS.


\end{document}